# Topological charge density around static colour sources in lattice QCD with dynamical quarks[*][†]

Manfried Faber, Harald Markum, Štefan Olejník[‡] and Wolfgang Sakuler

Institut für Kernphysik, Technische Universität Wien, A–1040 Vienna, Austria

We update our numerical investigation of topological structures around static quarks in pure gauge QCD by results of the first runs including dynamical quarks. Simulations were performed on an $8^3 \times 4$ lattice, with $SU(3)$ Wilson action, with 3 flavours of quarks of equal mass, both in the confinement and deconfinement phase. In the confinement phase we observe indications for the existence of a flux tube between a static quark and antiquark, flux-tube breaking for large separations, and local correlation between the topological charge density and chiral condensate. In the deconfinement phase almost all configurations turn out to be topologically trivial.

**1.** The *LATTICE 93* contribution of our group [1] presented results of the study of topological properties of QCD vacuum configurations with static colour sources in pure $SU(3)$ lattice gauge theory. The present work extends that investigation to $SU(3)$ with dynamical quarks. The main results from pure gauge QCD simulations can be summarized as follows [1,2]:

– Configurations with nonvanishing topological charges appear very seldom in the deconfinement phase.
– The topological charge density is lowered in the vicinity of external colour sources in both phases of pure gauge QCD.
– The suppression of the topological charge density appears in the whole flux tube between the static quark and antiquark in the confinement phase, while in the deconfinement phase the effect is concentrated around the (anti)quarks only.

It is of direct interest to verify if the above picture remains unchanged after taking dynamical quark degrees of freedom into account. In general one expects that most qualitative features of the confinement mechanism are the same in QCD without and with dynamical quarks. However, in a theory with fermions the topological susceptibility is proportional to a power of the quark mass [3]. Therefore, one can reasonably expect that the observed topological effects will depend on the presence of dynamical quarks and their masses.

**2.** Our calculations were performed on an $8^3 \times 4$ lattice with (anti)periodic boundary conditions using the Metropolis algorithm. We simulated QCD with $SU(3)$ gauge group, with standard Wilson action for gauge fields, and with 3 flavours of Kogut–Susskind quarks of equal mass, $ma = 0.1$. We performed runs for two values of $\beta$, one for the confinement ($\beta = 5.2$) and the other for the deconfinement phase ($\beta = 5.4$). In the former simulation observables of interest were measured on 2000 configurations separated by 50 sweeps, in the latter run only 300 configurations have been measured until now. The parameters of our runs were chosen in such a way so that the corresponding physical length scales were approximately the same as in our pure gauge simulations [1,2], thus enabling direct comparison of observed physical effects.

Topological properties of lattice configurations were probed by two local lattice operators of topological charge density:

$$q^{(P,H)}(x) = -\frac{1}{2^4 32\pi^2} \sum_{\mu,\ldots=\pm 1}^{\pm 4} \tilde{\epsilon}_{\mu\nu\rho\sigma} \mathrm{Tr}\, \mathcal{O}^{(P,H)}_{\mu\nu\rho\sigma}, \quad (1)$$

where the operators $\mathcal{O}^{(P,H)}_{\mu\nu\rho\sigma}$ are in usual notation

$$\mathcal{O}^{(P)}_{\mu\nu\rho\sigma} = U_{\mu\nu}(x) U_{\rho\sigma}(x), \qquad (2)$$

---

[*]Talk presented by Š. Olejník at *LATTICE 94*, Bielefeld, FRG, September 27–October 1, 1994.
[†]Supported in part by BMWF.
[‡]On leave from Institute of Physics, Slovak Academy of Sciences, SK–842 28 Bratislava, Slovak Republic.

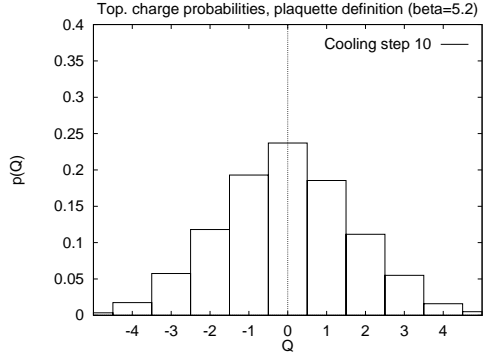

Figure 1. Probabilities for various values of the topological charge at $\beta = 5.2$.

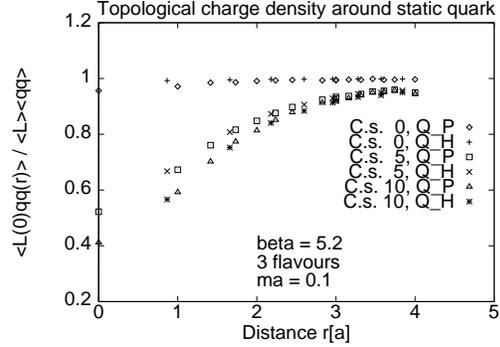

Figure 2. Squared topological charge density around a static quark at $\beta = 5.2$.

$$\begin{aligned}
\mathcal{O}^{(H)}_{\mu\nu\rho\sigma} &= U_\mu(x) U_\nu(x+\hat\mu) U_\rho(x+\hat\mu+\hat\nu) \\
&\times U_\sigma(x+\hat\mu+\hat\nu+\hat\rho) U^\dagger_\mu(x+\hat\nu+\hat\rho+\hat\sigma) \\
&\times U^\dagger_\nu(x+\hat\rho+\hat\sigma) U^\dagger_\rho(x+\hat\sigma) U^\dagger_\sigma(x) \quad (3)
\end{aligned}$$

(the "plaquette" ($P$) and the "hypercube" ($H$) operator [4]).

These operators are known to differ from the continuum topological charge density by multiplicative renormalization factors substantially different from unity at the values of $\beta$ used [5]. These factors were removed using the cooling method, in particular a slightly modified variant of Hoek et al. [6].

The local distribution of the topological charge density around single quarks and around a meson with quark-antiquark separation $d$ was obtained by calculating the correlation functions $\langle L(0) q^2(r) \rangle$ and $\langle L(0) L^\dagger(d) q^2(r) \rangle$, respectively, and normalizing them to the corresponding vacuum expectation values. $L(r)$ as usual denotes the Polyakov loop describing the propagation of a static colour charge.

3. The results of the simulation of full QCD at $\beta = 5.2$ are very similar to those of pure-gauge calculations at $\beta = 5.6$ [2]. After a few cooling steps topological charges of lattice configurations measured using the plaquette and hypercube operators coincide and reach nearly integer values. The probability distribution of the charges $Q$ is shown in Fig. 1. It is approximately Gaussian as in the pure-gauge case (cf. Fig. 2 of Ref. [2]).

The topological charge density (squared) is suppressed in the vicinity of a quark (Fig. 2). The suppression again occurs in the whole flux tube between the quark and antiquark (Fig. 3a,b). However, Fig. 3c reveals an essential difference between pure-gauge configurations and those containing dynamical quarks. At the largest $Q\bar Q$ separation, $d = 4$, the suppression is weaker in the middle of the tube than in the pure-gauge simulation. This we consider as a sign of flux-tube breaking due to the creation of a virtual dynamical $q\bar q$ pair. A similar effect was reported earlier in Ref. [7].

Our data also show correlations between the topological charge and the chiral condensate. The confinement configurations with higher topological charges tend to have higher values of the chiral condensate. The correlation appears also locally which is clearly seen from comparing Fig. 2 with Fig. 4, depicting the local chiral condensate distribution around a static quark.

The deconfinement simulations ($\beta = 5.4$) are still in progress. Most configurations belong

Topological charge density in a static meson

$\beta = 5.2$, $n_f = 3$, $ma = 0.1$

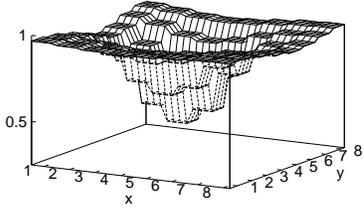

(a) 5 cooling steps, d = 2

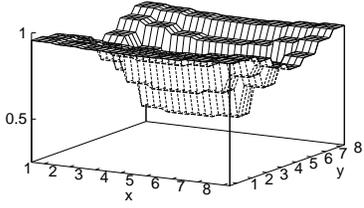

(b) 5 cooling steps, d = 3

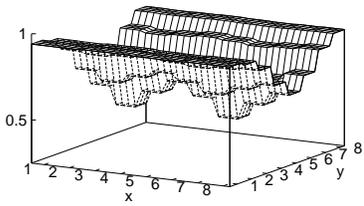

(c) 5 cooling steps, d = 4

Figure 3. Squared topological charge density around a static quark-antiquark pair at $\beta = 5.2$.

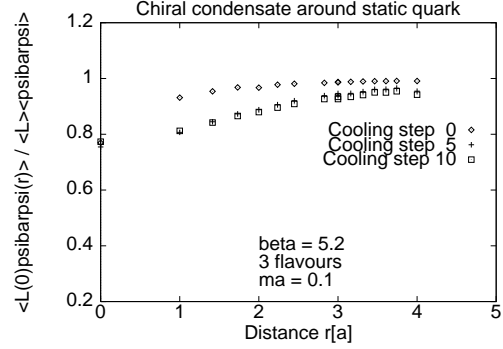

Figure 4. Local distribution of the chiral condensate around a static quark at $\beta = 5.2$.

to the topologically trivial sector, and non-zero charges appear even more rarely than in the pure-gauge case.

**4.** Our results from first runs with dynamical quarks indicate similar behaviour of topological quantities in pure-gauge QCD and full QCD, and hence a similarity in the corresponding confinement mechanisms. The quark-mass value used, however, was rather high and our conclusions have to be supported by simulations at lower mass values.